\def\year{2022}\relax
\documentclass[letterpaper]{article} 
\usepackage{aaai22}  
\usepackage{times}  
\usepackage{helvet}  
\usepackage{courier}  
\usepackage[hyphens]{url}  
\usepackage{graphicx} 
\usepackage{natbib}  
\usepackage{caption} 
\DeclareCaptionStyle{ruled}{labelfont=normalfont,labelsep=colon,strut=off} 
\usepackage{algorithm}
\usepackage{algorithmic}

\usepackage{newfloat}
\usepackage{listings}
\floatstyle{ruled}
\newfloat{listing}{tb}{lst}{}
\floatname{listing}{Listing}
\usepackage{xcolor}

\usepackage{booktabs}
\usepackage{colortbl}
\usepackage{threeparttable}
\usepackage{subcaption}
\title{Investigations of Performance and Bias in Human-AI Teamwork in Hiring}
\author {
    Andi Peng, \textsuperscript{\rm 1}$\thanks{Work done while an AI Resident at Microsoft Research.}$
    Besmira Nushi, \textsuperscript{\rm 2}
    Emre K\i c\i man, \textsuperscript{\rm 2}
    Kori Inkpen, \textsuperscript{\rm 2}
    Ece Kamar \textsuperscript{\rm 2}
}
\affiliations {
    \textsuperscript{\rm 1} Massachusetts Institute of Technology\\
    \textsuperscript{\rm 2} Microsoft Research\\
    andipeng@mit.edu, \{benushi, emrek, kori, eckamar\}@microsoft.com
}

\begin{document}

\maketitle

\begin{abstract}
In AI-assisted decision-making, effective \textit{hybrid} (human-AI) teamwork is not solely dependent on AI performance alone, but also on its impact on human decision-making. While prior work studies the effects of model accuracy on humans, we endeavour here to investigate the complex dynamics of how both a model's predictive performance and bias may transfer to humans in a recommendation-aided decision task. We consider the domain of ML-assisted hiring, where humans---operating in a constrained selection setting---can choose whether they wish to utilize a trained model's inferences to help select candidates from written biographies. We conduct a large-scale user study leveraging a re-created dataset of real bios from prior work, where humans predict the ground truth occupation of given candidates with and without the help of three different NLP classifiers (\textit{random, bag-of-words, and deep neural network)}. Our results demonstrate that while high-performance models significantly improve human performance in a hybrid setting, some models mitigate hybrid bias while others accentuate it. We examine these findings through the lens of decision conformity and observe that our model architecture choices have an impact on human-AI conformity and bias, motivating the explicit need to assess these complex dynamics prior to deployment.
\end{abstract}

\section{Introduction}
As AI-powered decision tools are increasingly deployed in real-world domains, a central challenge remains understanding how best to design models to assist humans~\cite{kleinberg2018human}. Ergo, a growing body of literature has arisen to study these \textit{screening} or \textit{recommendation} systems~\cite{kleinberg2019discrimination}, where a ML model acts as a data filtering mechanism to provide inferences as recommendations for a human decision-maker~\cite{gillies2016human}. These collaborative settings call for a different evaluation process prior. If the model were to operate alone, the typical evaluation pipeline would involve measuring and reporting various predictive performance metrics (i.e. \emph{how accurate is the model in solving the task?}), as well as checks for potential biases that may favor or disfavor groups based on sensitive attributes such as gender, age, or ethnicity (i.e. \emph{does the model exhibit lower predictive performance for a given group?}) ~\cite{mehrabi2021survey,barocas2017fairness}. Both axes (\emph{performance} and \emph{bias}) are important for real-world deployment and exhibit different social implications in practice~\cite{barocas2017fairness}. 


If the AI is instead intended to \textit{assist} the human rather than act as sole arbiter, then assessing resulting performance involves understanding the interaction between human and machine. When a human makes a decision with the help of an AI recommendation, they can either bring in their own perspectives in choosing how to utilize the model or may choose to solve the task alone. Thus, \textit{hybrid} (human-AI) performance depends on how the model alters the human decision, requiring an evaluation of a different nature that looks at how humans choose to conform to specific models.

\begin{figure}[t]
  \centering
  \includegraphics[width=\linewidth]{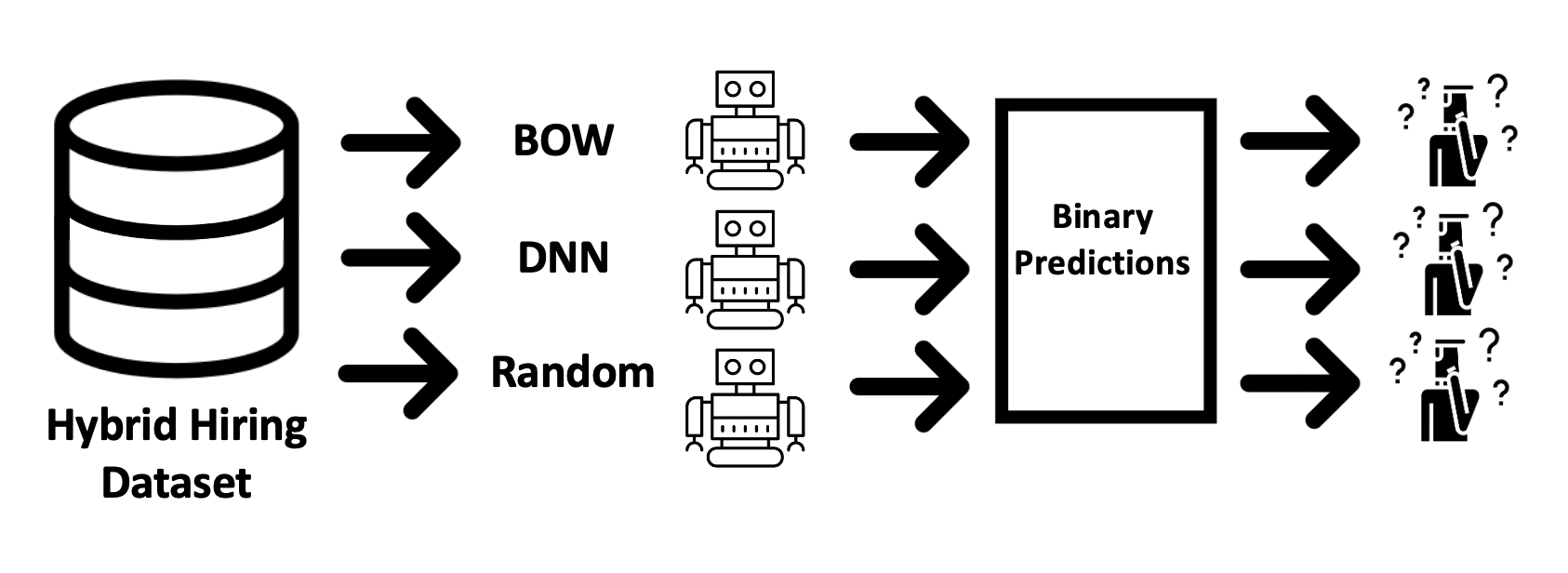}
  \caption{An example hybrid hiring workflow. A candidate dataset is used to train three NLP classifiers, which each outputs recommendations to human decision-makers. We evaluate accuracy and bias of the resulting system.}
  \label{task_design}
\end{figure}

Previous work has taken this approach in investigating how model accuracy transfers to hybrid accuracy \cite{lai-fat19,bansal2019beyond,green-cscw2019,feng-iui19}, illustrating that although hybrid systems designed for collaboration can improve accuracy beyond that of the human or AI alone, high model accuracy does not always transfer into high hybrid accuracy \cite{yin2019understanding}. However, despite this increasing focus on human-AI collaboration, the way predictive bias inherent in ML models transfer to human decisions is not well understood at all. Specifically, it is not clear how biases from different model architectures would influence human bias or whether a more biased model would ultimately propagate to a human decision-maker at a higher rate than a less biased one like in the case of accuracy. The two in combination (predictive performance and bias) result in complex dynamics that may alter how both percolate down to a human decision-maker.

In this work, we investigate this by conducting a large-scale study to assess how a realistic hybrid system performs on both overall accuracy and bias (difference in predicting male vs. female candidates). We choose the domain of hiring due to a rich literature of human and algorithmic biases documented, with the question at play being: ``Do I think this candidate is a good fit for this job?'' Our human study leverages a large-scale text dataset~\cite{de-arteaga2019bias} consisting of real candidate bios and employs three different NLP classifiers as assistance in predicting occupation from bio. We test how these models perform in isolation vs. when utilized as recommendations by humans in a hybrid system. To minimize side effects from other system properties (e.g., UX experience, confidence, etc.) we keep the interface presentation \textit{unchanged} in all conditions and display only the final model recommendation as an aid. Figure~\ref{task_design} illustrates our hybrid experimental setup.

We make the following contributions:
\begin{enumerate}
    \item To our knowledge, we present the first-ever experiment studying the propagation of both algorithmic performance and bias to human decision-making.
    \item Our results reveal surprising findings, demonstrating that some of our deployed models mitigate hybrid bias while others propagate and increase bias (even though original human and model biases span different regions). We interpret these results from a human-AI conformity lens and observe that high predictive performance from some model types do not necessarily increase human-model conformity, resulting in lower hybrid performance but less biased decisions. 
    \item We introduce our full crowdsourced data, comprised of 38,400 individual human judgements over 9,600 prediction tasks, as {\fontfamily{cmvtt}\selectfont {Hybrid Hiring}}: a first-ever large-scale dataset for studying human-AI collaborative decision-making trained, collected, and evaluated on real data.
\end{enumerate}

The above contributions provide important insights previously under-studied in both human-AI collaboration and algorithmic fairness literatures, and raise critical concerns and trade-offs that need to be investigated prior to deploying similar models in practice, particularly since our work revealed significant differences in model conformity, even \textit{without an interface change}. Inspired by these results, we propose future directions in studying the impact of different ML models in hybrid decision-making scenarios.

\section{Related Work}
\subsubsection{Algorithmic Bias}
It is now more important than ever to quantify and understand model biases that reinforce the disadvantaged status of different groups \cite{nosek2002harvesting,sweeney2013discrimination}. While ML achieves higher-still accuracy, a key question becomes: \textit{accurate, but for whom} \cite{barocas2016big}? Hiring, long a discriminatory practice \cite{isaac2009interventions}, has received specific renewed interest due to a rise in automated decision systems deployed with alarmingly detrimental effects towards female candidates \cite{amazon2018,raghavan2020mitigating}.

Spurred by such concerns, the ML community has responded with a rapidly growing body of literature on algorithmic fairness efforts. A brief overview ranges from approaches that seek to mitigate bias using techniques that are ``unaware'' of protected attributes like race and gender \cite{dwork2012fairness} to more sophisticated techniques that seek to impose fairness as a "constraint" \cite{hardt2016equality}. In practice, any method that relies on protected attributes for model training stands at odds with anti-discrimination law, which forbids the usage of these features in model prediction, even if the purpose is to mitigate bias \cite{dwork2018group,gonen2019lipstick}.

\subsubsection{Human Bias}
Complex decision tasks, limited cognitive resources, uncertain information, and a human tendency to aspire to reduce overall decision load together lead to a \textit{bounded rationality} model of human decision-making, where cognitive biases come into play \cite{simon1955behavioral,cunningham2013biases,kahneman2003aperspective}. These biases are best described as heuristics, or mental shortcuts, that humans take when evaluating large amounts of uncertain information in a messy world~\cite{thaler2008nudge}. One particular form of bias that has been found to be especially detrimental is that of gender bias, particularly when evaluating candidates in professional settings. There is evidence that gender inequalities in the workplace stem, at least in part, from biased attitudes directed against women from those who hold sexist or innate preferences for a particular gender in different professions \cite{koch2015meta}. For instance, a study found that the higher a participant scored on a hostile sexism personality test, the more likely they were to recommend a male candidate rather than female for a managerial position \cite{masser2004reinforcing}.

\subsubsection{Human-AI Collaboration}
The concept that decision processes adapt over time to adjust to changing preferences has led to \textit{preference construction}, or decision-makers formalizing which option they prefer \cite{lichtenstein2006construction,thaler2008nudge}. It is of no surprise that systems designed to produce recommendations in key stages of decision-making have been found to have immense impact on final outcomes \cite{mandl2011consumer}. In these cases, the human makes a decision to either \textit{accept} or \textit{reject} recommendations. These AI-assisted systems have led to more accurate medical diagnoses \cite{lundgard2018explainable}, optimized crowdsourcing efforts \cite{kamar2012combining}, and creative multiagent game-playing \cite{jaderberg2019human}. Here, we refer to human-AI together as a \textit{hybrid} system.

\begin{figure*}[t]
\centering
 \subcaptionbox{Human-only condition.}{\includegraphics[width=0.45\textwidth]{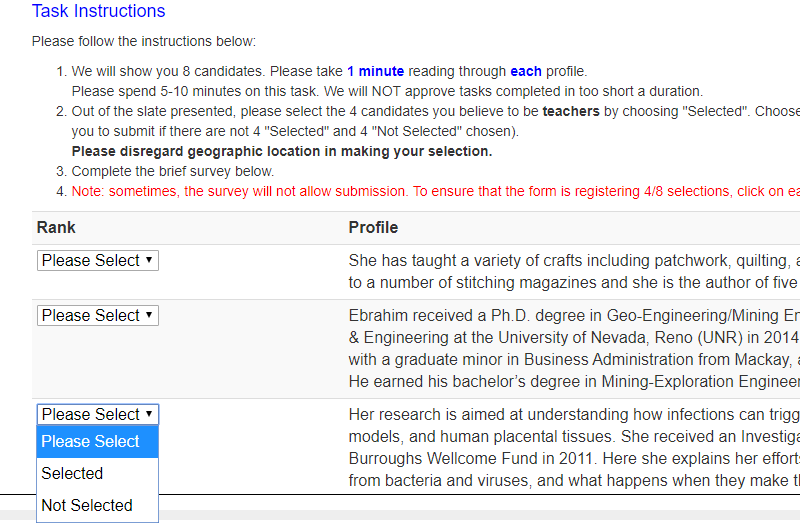}}
 \hfill
 \subcaptionbox{Hybrid condition.}{\includegraphics[width=0.45\textwidth]{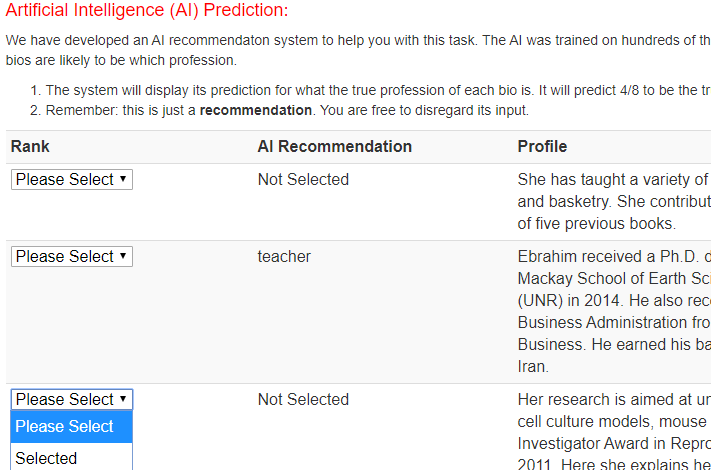}}
 \caption{An example task where the true occupation is \textit{teacher} and confused occupation \textit{professor}. The \textit{interface remains unchanged} across all candidate slates and conditions. Additionally, recommendations do not provide any additional evidence or signal of the underlying model behaviour (e.g. confidence, architecture, explanation for the decision, etc.).}
  \label{fig:task}
\end{figure*}

As hybrid systems are increasingly deployed, it is important to understand their impact on human decision processes. Many factors, such as the human's ability to create a mental picture of the model \cite{bansal2019beyond}, their implicit trust in the model \cite{yin2019understanding,zhang2020effect}, how they are impacted by updates \cite{bansal2019updates}, the representational display of recommendations \cite{peng2019what}, and the interpretability of the model \cite{gilpin2018explaining} have all been demonstrated to greatly impact humans. However, to our knowledge, there exists no work that studies how both AI predictive performance and bias transfer to humans.

\section{Experimental Setup}
\subsubsection{Motivation}
Our work is motivated by the desire to understand how bias in algorithmic models transfer to hybrid decision-making in realistic deployed settings where both users of trained models and their real-world stakeholders are impacted. Often, it is assumed that a higher-performing model will help a human make more-accurate and less-biased decisions, or conversely, that a human will recognize model mistakes and exert agency in correcting them. Yet, we have very little understanding of how these metrics trickle down through a hybrid decision pipeline. In this work, we evaluate how different models trained on real-world data, when integrated within common hiring pipeline under constraints, alter final system predictive performance and bias. Studying this allows us to better understand the impact of this increasingly-common workflow as well as unearth which types of algorithmic advancements can actually be transferred to a human-in-the-loop system.

\subsubsection{Data Collection}
We select the task of language-based \textit{occupation classification} due to its direct relevance to real-world hiring scenarios \cite{peng2019what}. To a human, predicting an individual's true occupation from a brief text description remains a common and often high-stakes decision made in professional settings daily. We compile a corpus of public professional bios using the same methodology as De-Arteaga et al. by scraping online bios using the Common Crawl to re-create a dataset where all observations begin with the following sequence: [\textit{name} is a \textit{title}] and subsequently describe a professional background \cite{de-arteaga2019bias}. We extract the ground truth occupation and gender of each observation and to the best of our ability, mask out names. We select the 28 most frequently-occurring occupations, resulting in 397,907 observations of which \textit{professor} is the most frequent occupation and \textit{rapper} the least. This dataset represents a publicly-available online pool of candidates that may be screened by a real model.

\subsubsection{Model Training}
The objective is to, without access to the first sentence of a bio which identifies occupation, predict the ground truth using the candidate's self-provided description. To isolate the impact of model architecture on hybrid performance, we elect to train a single-layer fully-connected \textit{deep neural network} (DNN) as well as a simpler \textit{bag of words} (BOW) \cite{de-arteaga2019bias,bolukbasi2016man}. For our BOW, we use a one-versus-all logistic regression with L$_{2}$ regularization similar with prior work \cite{de-arteaga2019bias,romanov2019what}. DNN represents a more \textit{black-box} architecture due to its non-linear nature and deeply nested structures whereas BOW remains a good baseline due to its general \textit{interpretability} \cite{gilpin2018explaining}.

\begin{figure*}[t]
\centering
 \subcaptionbox{BOW classification bias.}{\includegraphics[width=0.45\textwidth]{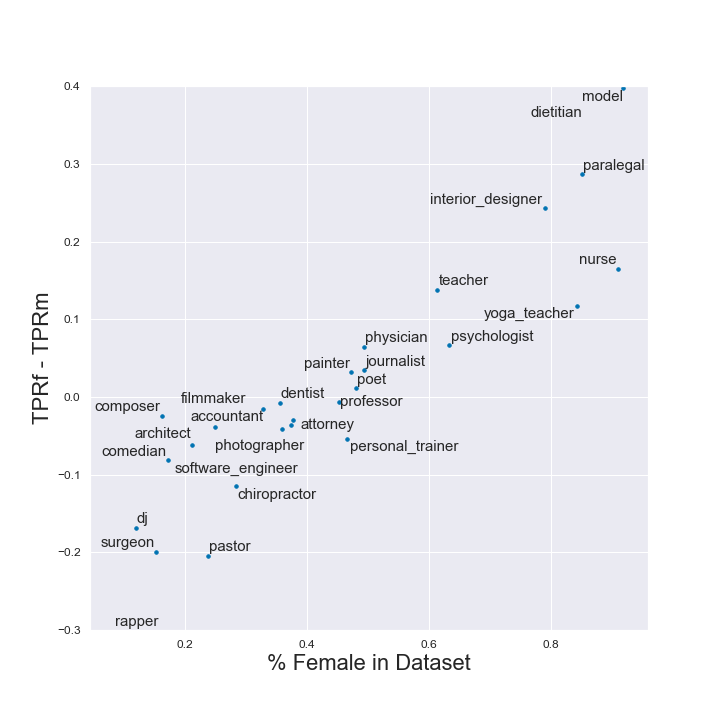}}
 \hfill
 \subcaptionbox{DNN classification bias.}{\includegraphics[width=0.45\textwidth]{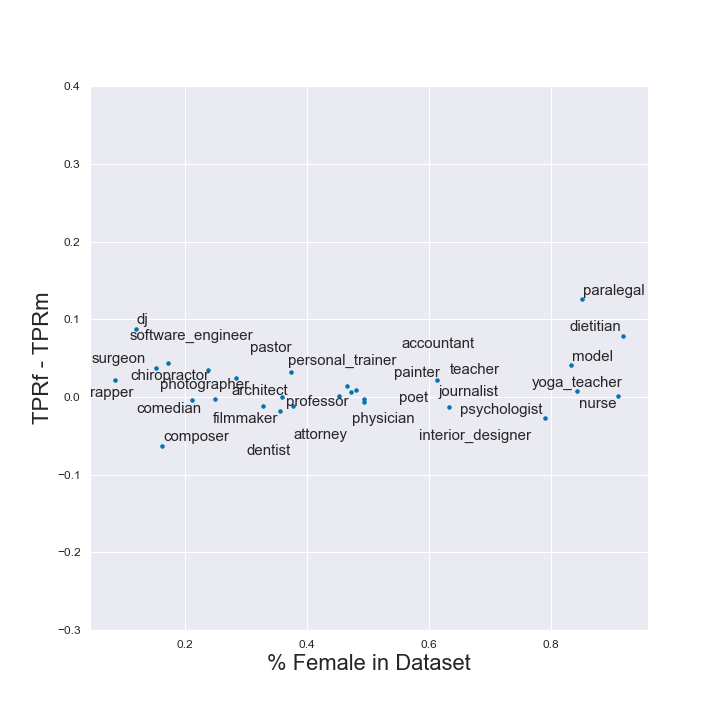}}
 \caption{DNN and BOW gender bias on the dataset test split as quantified by TPR gender gap ($\Delta$TPR) relative to true proportion of female candidates in the dataset. While both models exhibit biases, DNN's $\Delta$TPRs across occupations do not appear as extreme as BOW's. Note that our candidate slates are generated from bios sampled from this distribution.}
  \label{fig:model_bias}
\end{figure*}

Because some occupations exhibit an uneven skew of either male or female bios and we wish to de-link existing data pipeline biases from our analysis, we create validation and test splits such that both gender and occupation are sufficiently represented. In accordance with prior work \cite{de-arteaga2019bias,romanov2019what}, we use stratified-by-occupation splits, with 65\% of the bios (258,639) designated for training, 10\% (39,790 bios) designated for validation, and 25\% (99,476 bios) designated for testing. This isolates the differences in model performance to their varying architectures, and allows for an equivalent apples-to-apples comparison on resulting hybrid performance and bias.

\subsubsection{Human Task Design}
We construct a constrained decision task by presenting 8 bios, 4 of which belong to the occupation of interest and ask humans to identify the correct 4 out of the 8 that belongs to that occupation. We are in effect simulating a realistic scenario where, say, a recruiter operating under resource constraints is tasked with selecting a subset of candidates for interviewing and may make implicit judgements based on gender \cite{amazon2018}. To ensure that our slates are non-trivially difficult for humans, we generate confusion matrices for predictions made by our models and select the following 3 pairs of highly-confused professions by gender: \textit{attorney} and \textit{paralegal}, \textit{surgeon} and \textit{physician}, and \textit{professor} and \textit{teacher}. Then, to assess the potentially bi-directional nature of bias (for example, a female lawyer being misclassified as a paralegal implies something very different than a male paralegal being misclassified as a lawyer), we create 6 tasks from these 3 occupation pairs (i.e. one type of slate is an attorney misclassified as a paralegal and its counterpart a paralegal misclassified as an attorney).

For each occupation, we design candidate slates where 8 bios are randomly selected from our test split (4 from the true occupation and 4 from the confused occupation), with the additional constraint that gender representation remain equal in both. This is done to enforce the opportunity to select equal subsets of ``qualified'' candidates, irrespective of how they are actually represented in the world. Altogether, we generate 200 unique slates, randomly ordered, for each occupation to total 9,600 samples from our original dataset (6*200*8 = 9,600 bios total to be classified by each control group).

\subsubsection{Evaluation}
To study the impact of AI recommendations on human decision-making, we conduct a crowdsourced study across three conditions (model-only, human-only, and hybrid) and evaluate the following two metrics:
\begin{enumerate}
    \item Predictive performance (true positive rate (TPR))
    \item Bias (differential TPR in classifying female vs. male candidates ($\Delta$TPR, or TPR$_{f}$ - TPR$_{m}$))
\end{enumerate}
Note, these two axes are not the same: a system may classify candidates successfully at a higher rate but also exhibit bias in being more accurate for male vs. female candidates. The ideal system is one that maximizes TPR without exhibiting significant $\Delta$TPR. We report TPR rather than accuracy since we are studying constrained decision-making where the candidate slate size is fixed and if one classification is correct, this necessitates that another was incorrect. This helps our evaluation of bias ($\Delta$TPR), which is calculated as the difference in TPRs between binary gender candidates of each occupation \cite{de-arteaga2019bias}. A positive $\Delta$TPR indicates a bias towards female candidates and negative $\Delta$TPR) towards male. In line with previous work \cite{peng2019what}, we formulate the task as a filtering rather than a classification task, which allows for us to observe bias to a greater extent since a budget is allocated for selection and not all candidates can be prioritized (as is the case in real-world settings). A biased system will exhibit a statistically significant $\Delta$TPR (i.e. TPR$_{f}$ $\neq$ TPR$_{m}$) across slates.

\begin{table*}
 \centering
 \renewcommand*\TPTnoteLabel[1]{\parbox[b]{3em}{\hfill#1\,}}
 \begin{threeparttable}
  \caption{TPR on the same candidates slates across conditions. Pairwise comparisons are made between the human (base condition) and each corresponding model to assess the performance differential. Higher TPR models (DNN and BOW) consistently translate into higher TPR hybrid systems (H+DNN and H+BOW) whereas a lower TPR model (Random) does not impede performance (H+R).}
  \begin{tabular}{lccccccc}\toprule
      & \textbf{Human} & \textbf{Rand} & \textbf{H+R} & \textbf{DNN} & \textbf{H+DNN} & \textbf{BOW} & \textbf{H+BOW}\\\midrule
     attorney & 0.60 & \cellcolor{green}$0.51^\beta$ & 0.57 & \cellcolor{yellow}$0.79^\alpha$ & \cellcolor{yellow}$0.66^\alpha$ & \cellcolor{yellow}$0.78^\alpha$ & \cellcolor{yellow}$0.70^\alpha$\\
     paralegal & 0.60 & \cellcolor{green}$0.49^\beta$ & 0.56 & \cellcolor{yellow}$0.87^\alpha$ & \cellcolor{yellow}$0.68^\alpha$ & \cellcolor{yellow}$0.78^\alpha$ & \cellcolor{yellow}$0.70^\alpha$\\
     physician & 0.52 & \cellcolor{green}$0.49^\beta$ & 0.52 & \cellcolor{yellow}$0.85^\alpha$ & \cellcolor{yellow}$0.61^\alpha$ & \cellcolor{yellow}$0.85^\alpha$ & \cellcolor{yellow}$0.66^\alpha$\\
     surgeon & 0.61 & \cellcolor{green}$0.51^\beta$ & 0.61 & \cellcolor{yellow}$0.89^\alpha$ & \cellcolor{yellow}$0.68^\alpha$ & \cellcolor{yellow}$0.82^\alpha$ & \cellcolor{yellow}$0.74^\alpha$\\
     professor & 0.59 & \cellcolor{green}$0.51^\beta$ & 0.59 & \cellcolor{yellow}$0.85^\alpha$ & \cellcolor{yellow}$0.70^\alpha$ & \cellcolor{yellow}$0.87^\alpha$ & \cellcolor{yellow}$0.75^\alpha$\\
     teacher & 0.53 & \cellcolor{green}$0.50^\beta$ & 0.54 & \cellcolor{yellow}$0.86^\alpha$ & \cellcolor{yellow}$0.61^\alpha$ & \cellcolor{yellow}$0.87^\alpha$ & \cellcolor{yellow}$0.74^\alpha$\\
  \bottomrule
  \end{tabular}
 \footnotesize
  \begin{tablenotes}
   \item[\tnote{$\alpha$}] Greater than the Human condition, significant at $p<0.01$. Also in yellow.
   \item[\tnote{$\beta$}] Less than the Human condition, significant at $p<0.01$. Also in green.
  \end{tablenotes}
  \label{tab:tpr}
 \end{threeparttable}
\end{table*}

\subsubsection{Model-Only Condition}
For each of our generated candidate slates, AI recommendations are created by selecting the top 4 bios that our trained DNN and BOW models have the highest confidence in their predictions as belonging to the ground truth occupation. This forces the same constrained decision task that our subsequent conditions will face. In addition, we also test a ``random'' model, which selects its 4 bios via coin flip to serve as a non-intelligent baseline. Because we are enforcing the same subset criteria on the exact same candidate slates, we can attribute any arising performance differences to model type and not the task itself.

\subsubsection{Human-Only Condition}
For our human-only condition, we deploy slates as HIT tasks on mTurk (Figure~\ref{fig:model_bias}). We show each participant a unique slate, present a description of the ground truth occupation, and ask them to select 4/8 bios that they believe to best fit that description. We programmatically enforce that each participant picks the correct number of selections and each bio must be user-clicked as \textit{Selected} or \textit{Not Selected}. Bios are randomly ordered per slate to remove possible confounding factors such as rank ordering preference and recency bias (although final generated slates are kept consistent between conditions). Altogether, we deploy 1,600 uniquely-generated HITs across six tested occupations.

\subsubsection{Hybrid Condition}
For our hybrid condition, we follow the same methodology as for our human-only condition but additionally provide predictions made by our three models. Participants are explicitly instructed that these predictions are ``recommendations'' from an ``AI'' that they may choose to disregard and override. For this condition, we deploy 4,800 unique HITs in total (1,600 each for human+DNN, human+BOW, and human+Random). Note: irrespective of the model tested, the interface remained the same and participants could not participate in HITs across conditions.

To increase reproducibility confidence, we run all 200 slates per occupation in two batches of 100 across unique study participant pools, each with a mix of human-only and hybrid conditions: the first between August 23-27, 2019 and the second between September 1-4, 2019. This is done to ensure that demographic skews in crowdsourcing may be mitigated across worker pools. We compensate all participants at a wage of \$15 per hour. Participants are additionally screened according to the following qualifications: hold above a 95\% approval rating, unique ID per condition, and based in the United States to control for English being the primary spoken language.

\subsubsection{Data Ethics and Privacy}
For all experiments and collected data, we conduct both institutional IRB  and data privacy review. We also anonymize all bios (by stripping out names and other identifying features) and participant data (we collect no no personal or private information).

\subsubsection{Statistical Testing}
In evaluating significance across conditions, we are interested in seeing whether a condition (i.e. a specific model) produces changes in hybrid performance when compared to a baseline. We use the human-only condition as our baseline for all comparisons since we are interested in studying the impacts of AI on humans in this work. We utilize Friedman and Wilcoxon signed ranks tests to study the effect of each candidate slate across conditions in pairwise comparisons to the human-only (base) condition.

\begin{table*}[t]
 \centering
 \renewcommand*\TPTnoteLabel[1]{\parbox[b]{3em}{\hfill#1\,}}
 \begin{threeparttable}
  \caption{Bias ($\Delta$TPR) across conditions for tested occupations. Within each slate, we conduct a pairwise comparison between TPR$_{f}$ and TPR$_{m}$ to see whether a significant difference is present. If so, that condition exhibits a significant $\Delta$TPR.}
  \begin{tabular}{lccccccc}\toprule
      & \textbf{Human} & \textbf{Rand} & \textbf{H+R} & \textbf{DNN} & \textbf{H+DNN} & \textbf{BOW} & \textbf{H+BOW}\\\midrule
     attorney & -0.02 & -0.04 & -0.02 & -0.04 & -0.03 & -0.06 & -0.03\\
     paralegal & \cellcolor{pink}0.09\tnote{*} & 0.03 & 0.07 & \cellcolor{pink}0.11\tnote{*} & 0.03 & \cellcolor{pink}0.23\tnote{*} & \cellcolor{pink}0.15\tnote{*}\\
     physician & -0.02 & 0.02 & -0.00 & \cellcolor{pink}0.09\tnote{*} & -0.00 & 0.05 & 0.06\\
     surgeon & -0.06 & -0.04 & \cellcolor{pink}-0.13\tnote{*} & \cellcolor{pink}-0.07\tnote{*} & -0.03 & \cellcolor{pink}-0.16\tnote{*} & \cellcolor{pink}-0.16\tnote{*}\\
     professor & 0.02 & 0.04 & 0.00 & -0.04 & -0.03 & -0.06 & -0.03\\
     teacher & \cellcolor{pink}0.10\tnote{*} & -0.03 & 0.03 & 0.03 & 0.02 & 0.04 & 0.07\\
  \bottomrule
  \end{tabular}
 \footnotesize
  \begin{tablenotes}
   \item[\tnote{*}] TPR$_f \neq $TPR$_m$, significant at $p<0.01$. Also in pink.
  \end{tablenotes}
  \label{tab:bias}
 \end{threeparttable}
\end{table*}

\subsection{Results}
First, we examine performance of our model-only condition. We see that different models exhibit different TPRs and biases, with BOW and DNN architectures indeed making varied selections on the same task. Second, we turn to the human-only condition and find that humans exhibit their own set of biases that do not parallel either trained model. Third, we assess the impact of recommendations on human decision-making in our hybrid condition and find that although a higher-TPR model consistently produces higher-TPR hybrid teamwork, the impact on bias is model-specific, with DNN mitigating human bias while BOW seemingly inducing it. Last, we assess these results through the lens of human-AI conformity and discover that high-TPR performance from our tested non-linear model does not necessarily increase human-model agreement, resulting in ultimately lower hybrid performance but less biased decisions.

\subsubsection{Model-Only Performance}
Table~\ref{tab:tpr} highlights the TPRs of human and model-only conditions. We see that DNN and BOW do not make identical predictions across candidate slates, with DNN generally outperforming BOW (as evidenced by the difference in TPRs, particularly on \textit{paralegal} and \textit{surgeon} tasks). To probe this further, we analyze the original classifications made by both models and find that, as shown in Figure~\ref{fig:model_bias}, DNN and BOW exhibit different biases ($\Delta$TPRs) across occupations. For example, BOW $\Delta$TPR of \textit{paralegals} (top right of Figure 3a) indicates both a true high proportion of female paralegals in the dataset as well as model bias in classifying them as such.

\subsubsection{Human-Only Performance}
We next ask the question: do human predictions resemble that of either model? Across both TPR and $\Delta$TPR evaluations, we find that human-only decisions do not overlap with those from either BOW or DNN-only predictions at different rates, thus removing the possible confounder that one model aligned with original human decisions more than the other (details can be found in Appendix). Table~\ref{tab:tpr} shows that the human-only condition significantly under-performs both DNN and BOW models on all occupation slates, although in most cases does perform higher than Random. Moreover, Table~\ref{tab:bias} illustrates different biases across different conditions, with DNN not exhibiting any significant bias across all occupations, BOW biased towards female \textit{paralegals} and male \textit{surgeons}, and humans biased towards female \textit{paralegals} and \textit{teachers}.

\subsubsection{Model-Specific Impact On Hybrid TPR}
When assessing the impact of model TPR on hybrid decision-making, we find that human decision-makers collaborating with a higher TPR model (DNN and BOW) results in a consistently significant \textit{improvement} across all occupations. This is in accordance with previous work, which has observed that higher-accuracy models generally help lower-accuracy humans \cite{bansal2019beyond,bansal2021does}, although this is still far from achieving optimal complementarity. Interestingly, when humans collaborate with a lower TPR model (Random), their own performance is \textit{not impeded} (Table~\ref{tab:tpr}).

\begin{figure}[t]
  \centering
  \includegraphics[width=\linewidth]{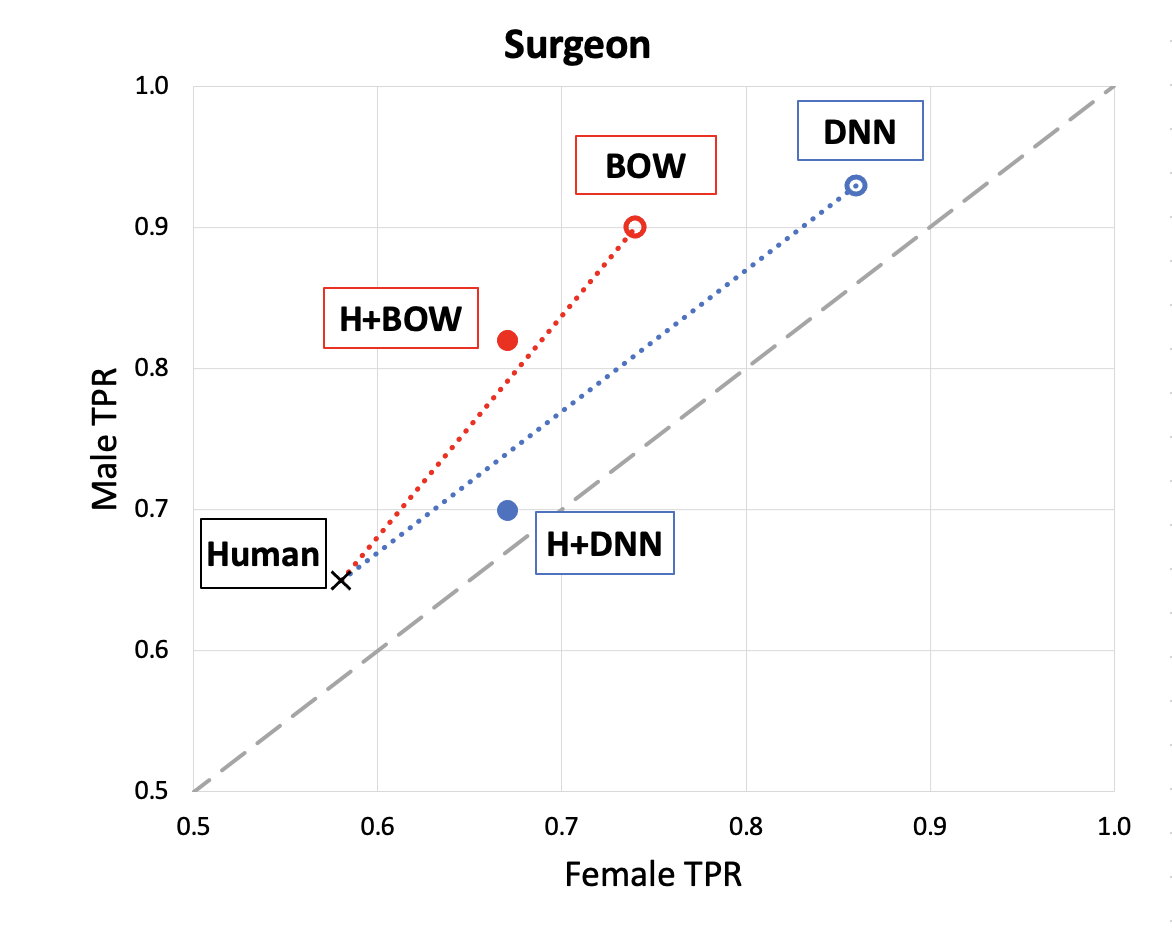}
  \caption{A visual of bias within the \textit{surgeon} task, plotted again female (x-axis) and male (y-axis) TPRs. The center (grey) line represents an unbiased model. The bottom left represents a less accurate model, and the top right more accurate. Interpolation (dotted) lines are drawn to represent the expected trendline if no consistent difference across hybrid conditions existed. We see that DNN helps mitigate human bias (the resulting hybrid $\Delta$TPR is close to the unbiased line) whereas BOW appears to actually induce bias (resulting in a hybrid $\Delta$TPR farther from the line).}
  \label{fig:surgeon}
\end{figure}

\subsubsection{Model-Specific Impact On Hybrid $\Delta$TPR}
A different story emerges when evaluating the impact of model $\Delta$TPR on hybrid decision-making, with different models impacting resulting biases differently. When humans collaborate with DNN, the resulting system (irrespective of any human biases at play) becomes unbiased. Table~\ref{tab:bias} illustrates how the originally biased occupations of \textit{paralegal} and \textit{teacher} become both mitigated by an unbiased DNN. However, an opposite effect can be seen in humans collaborating with BOW, with the resulting system seemingly reflecting both human-only and model-only biases. For example, despite the original human being unbiased in the \textit{surgeon} task, the resulting hybrid system is pulled towards a significant bias towards male candidates. Figure ~\ref{fig:surgeon} analyzes this result in greater detail using the \emph{surgeon} task as an illustration. Note that the key point is not only that the DNN-hybrid system is ultimately less biased than the BOW-hybrid (lower hybrid $\Delta$TPR), but that the resulting system is pulled below the interpolated expected (blue) line between Human and DNN performance gains towards the fully unbiased (grey) line, whereas the BOW-hybrid is pulled above the interpolated (red) line towards a more biased direction. Visually, this helps differentiate between bias mitigation that may result from performance gains of a higher-performing model and highlights differences between how bias percolates differently from a DNN vs. BOW model down to a human.

\subsubsection{Investigating Conformity}
\begin{figure}
  \centering
  \includegraphics[width=0.9\linewidth]{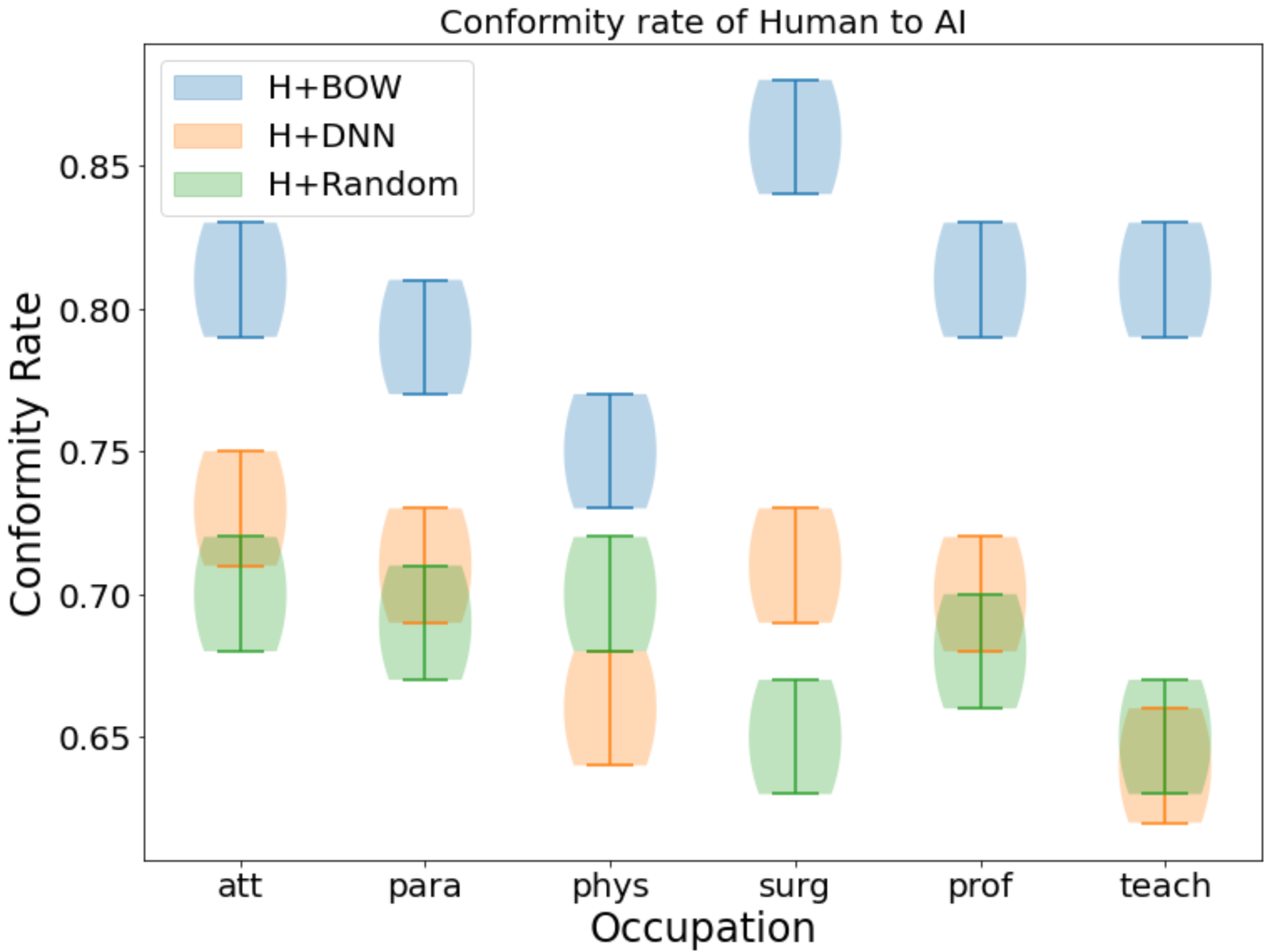}
  \caption{Conformity rate (percentage of hybrid decisions that match those predicted by the model alone) across tested occupations. We see significantly higher conformity to BOW than to DNN and Random predictions, with highlighted bands detailing 95\% confidence intervals.}
  \label{fig:conformity}
\end{figure}

Why do we see very different results for model-specific impacts on TPR vs. $\Delta$TPR hybrid decision-making, \textit{even without an interface change}? To better understand a sample-by-sample breakdown, we investigate human-AI conformity, or the rate at which a human appears to follow the model's recommendations in a hybrid system. We compute this by assessing the percentage of hybrid decisions that match those of original model decisions for each candidate slate (irrespective of whether that classification was the ground truth or not). Figure~\ref{fig:conformity} illustrates that although we see similar conformity rates of the human to DNN, humans conform significantly more to BOW predictions than either DNN or Random. Moreover, this distinction is especially apparent in cases where the model made an incorrect prediction (Table~\ref{tab:bad}). A possible explanation, supported by past work, posits that BOW is a generally more \textit{interpretable} model that humans can understand (and trust) more \cite{de-arteaga2019bias}. Because BOW word associations are learned by encoding sparse vectors that map to word vocabularies in a manner that is thought of to be more linear, humans are able to formulate an internal understanding of its recommendations more readily than DNN (a \textit{black-box} non-linear model) or Random (complete chance) \cite{bansal2019beyond,poursabzi-sangdeh2018manipulating}. In fact, based on Table~\ref{tab:bias} we observe that despite the lower hybrid performance of the Random model, random recommendations appear to have similar effects to the DNN on mitigating mitigating bias. As a result, humans may be more willing to accept the inferences provided by BOW (even when those recommendations are biased) and conform to its predictions, particularly when operating under resource constraints.

\begin{table}[t]
 \centering
 \caption{Hybrid decisions that match original model decisions, conditioned on the model being incorrect, i.e. \textit{when does a human accept a wrong prediction?} Here, H+Random serves as a baseline for understanding the additional conformity to a specific architecture beyond blind acceptance of AI recommendations themselves. We observe that humans are significantly more likely to conform to incorrect BOW decisions relative to DNN, which rarely differs from Random.}
 \label{tab:bad}
 \renewcommand*\TPTnoteLabel[1]{\parbox[b]{3em}{\hfill#1\,}}
 \begin{threeparttable}
  \begin{tabular}{lccccccc}\toprule
      & \textbf{H+Random} & \textbf{H+DNN} & \textbf{H+BOW}\\\midrule
     attorney & 0.622 & 0.663 & \cellcolor{cyan}0.744\tnote{*} \\
     paralegal & 0.629 & 0.634 & \cellcolor{cyan}0.716\tnote{*} \\
     physician & 0.673 & 0.648 & \cellcolor{cyan}0.782\tnote{*} \\
     surgeon & 0.561 & \cellcolor{cyan}0.645\tnote{*} & \cellcolor{cyan}0.809\tnote{*} \\
     professor & 0.605 & 0.504 & \cellcolor{cyan}0.704\tnote{*} \\
     teacher & 0.606 & 0.623 & \cellcolor{cyan}0.804\tnote{*} \\
  \bottomrule
  \end{tabular}
  \footnotesize
  \begin{tablenotes}
   \item[\tnote{*}] Greater than H+Random when the model is incorrect, significant at $p<0.01$. Also in blue.
  \end{tablenotes}
 \end{threeparttable}
\end{table}

\begin{table}
 \centering
 \caption{Prediction overlap between the human-only and model-only conditions, i.e. \textit{what percentage of the original human decisions matched those of each model?} Although we see higher human overlap with DNN and BOW vs. Random (likely due to Random being a generally lower-performing model that operates by chance), there is no significant difference between DNN vs. BOW. This helps assuage concerns regarding one model resembling human reasoning more than another prior to deployment in the task.}
 \renewcommand*\TPTnoteLabel[1]{\parbox[b]{3em}{\hfill#1\,}}
 \begin{threeparttable}
  \begin{tabular}{lccccccc}\toprule
      & \textbf{Random} & \textbf{DNN} & \textbf{BOW}\\\midrule
     attorney & 0.511 & \cellcolor{green}0.589\tnote{*} & \cellcolor{green}0.608\tnote{*} \\
     paralegal & 0.498 & \cellcolor{green}0.583\tnote{*} & \cellcolor{green}0.570\tnote{*} \\
     physician & 0.501 & 0.510 & \cellcolor{green}0.526\tnote{*} \\
     surgeon & 0.485 & \cellcolor{green}0.599\tnote{*} & \cellcolor{green}0.575\tnote{*} \\
     professor & 0.510 & \cellcolor{green}0.554\tnote{*} & \cellcolor{green}0.570\tnote{*} \\
     teacher & 0.516 & \cellcolor{green}0.531\tnote{*} & \cellcolor{green}0.526\tnote{*} \\
  \bottomrule
  \end{tabular}
  \footnotesize
  \begin{tablenotes}
   \item[\tnote{*}] Greater than Random, significant at $p<0.01$. Also in green.
  \end{tablenotes}
 \end{threeparttable}
\end{table}

\subsection{Discussion}

\subsubsection{Impact on Model Deployment}
A natural question that arises from these findings is whether DNN and Random (which both appear to be uninterpretable models) help mitigate human biases because they force human decision-makers to self-reflect more, and if so, whether ML deployment should actually prioritize this objective in future system design where minimizing bias may be a priority. To do so would mean an orthogonal departure from current work, where system designers are seeking less biased and more interpretable models. Moreover, our H+BOW was more accurate than our H+DNN, posing a trade-off between high team accuracy vs. low team bias. Our recommendation is that, while our results are somewhat surprising and highlight the importance of studying real-world hybrid decision-making, deploying a less interpretable model serves as a shortcut to true bias mitigation. As a community, we should seek to discover mechanisms that achieve this more explicitly and efficiently to truly leverage the complementary strengths of improved algorithmic design. Examples may include requiring humans to follow explicit forms of self-reflection and decision justification when there exists a a risk of bias.

\subsubsection{Dataset Release}
We introduce our full experimental data as {\fontfamily{cmvtt}\selectfont {Hybrid Hiring}}, a large-scale dataset for studying human-AI decision-making that is collected and evaluated on real-world candidates. Comprised of 38,400 human judgements over 9,600 unique prediction tasks across seven conditions, our dataset represents a first of its kind released to study human decision-making in the loop utilizing trained ML inferences. Ideally, hiring (and other high-stakes social decisions) should always remain in the purview of human review, and so utilizing datasets and methodologies of this kind will allow the field to investigate the impacts of different research questions on human decision-making in these contexts. Although we specifically investigated hybrid performance of three NLP models, one can easily extend this work to alternate architectures and interfaces.

\subsubsection{Limitations}
While we do our best to simulate a realistic hybrid task by selecting a socially relevant domain where real human data is incorporated in the decision-making of human study participants, we recognize that we are still running a controlled study on mTurk, where transfer of results to real-world deployed systems may be limited. Moreover, we greatly simplify many potential confounders (such as age, presence of non-binary gender, and self-written biography variance) in isolating bias to a single variable. We also do not study state-of-the-art de-biased models due to more complex architectures and leave for future work. We hope that our work moves the needle more in the direction of studying the impacts of ML-aided systems in real-world environments and propose that the community jointly invest in producing similar large-scale decision tasks and datasets to further study such intricacies across varied domains.

\subsection{Conclusion}
In asking the question of how model performance impacts human decision-making on two axes, our findings open up additional questions related to the specificity of human responses to different models, \textit{even without an interface change}. Our results motivate the explicit need to further investigate the observed signals regarding differing human intuitions of varied model architectures and how we can best design systems that allow for optimal hybrid collaboration.

\subsection{Acknowledgements}
We would like to thank Adam Kalai and Maria De-Arteaga for helpful discussions on problem formulation, Alexey Romanov for help with data collection and model training, Sarah Jobalia for moral support, and the anonymous reviewers for comments on the draft. Andi Peng is supported by an NSF Graduate Research Fellowship.

\bibliography{citations.bib}
\clearpage
\appendix

\end{document}